\documentclass[aps,prb,twocolumn,showpacs]{revtex4}

\usepackage{amsmath}
\usepackage{amsfonts}
\usepackage{amssymb}
\usepackage{graphicx}
\usepackage{bm}

\begin{document}

\title{\textit{Additional} carrier-mediated ferromagnetism in GdN}

\author{Anand Sharma}
\email{anand@physik.hu-berlin.de}
\altaffiliation[Present address: ]{NRC, IMS, M-50, 1200 Montreal Road, Ottawa, K1A0R6, ON, Canada.}

\affiliation{Institut f$\ddot{u}$r Physik, Humboldt-Universit$\ddot{a}$t zu Berlin, Newtonstr. 15, 12489, Berlin, Germany.}

\author{Wolfgang Nolting}
\affiliation{Institut f$\ddot{u}$r Physik, Humboldt-Universit$\ddot{a}$t zu Berlin, Newtonstr. 15, 12489, Berlin, Germany.}

\begin{abstract}
The mechanism behind ferromagnetic exchange interaction in GdN is not well understood. It has been argued that it can be due to fourth order cross process of \textit{d-f} mixing and \textit{d-f} exchange. An alternative explanation suggests an anti- ferromagnetic interaction between Gd \textit{d} and N \textit{p} induced moments on the rock salt structure which aligns the nearest neighbor Gd \textit{f} moments ferromagnetically through the \textit{d-f} exchange. In this paper we present results of Curie temperature in GdN as a function of carrier density calculated within our multiband modified RKKY- like exchange interaction. It includes realistic bandstructure of the 5\textit{d} conduction band as an input for single particle energies. We analyze the possibility of carrier- mediated ferromagnetism in GdN and also demonstrate a simple phenomenological model which justifies the role of charge carriers.
\end{abstract}

\date{\today}


\maketitle

\section{\label{sec:intro} Introduction}
\indent Spintronics~\cite{wolf,zutic} is a technology which utilizes the charge states of electrons as in a semiconductor as well as the quantum spin states as used in the data storage devices. The interest in semiconductor based spintronics has greatly accelerated the studies of magnetic semiconductors~\cite{matnol}. The aim is to find semiconductors exhibiting magnetism at or above room temperature. In order to achieve this a host semiconductor material was doped with transition metal ions (Mn) producing diluted magnetic semiconductors (DMS) with Curie temperature, T$_{\textrm{c}}$, as high as 185 K~\cite{novak}. Though there have been a lot of progress during last decades to accomplish this task but there is still lack of complete understanding due to complications in growth and measurement techniques and because of approximate theories.\\
\indent Among other class of materials are the strongly correlated rare earth (RE) compounds having incompletely filled \textit{f} electron shells. In particular, Gadolinium Nitride (GdN) has been the most widely studied~\cite{duan1} owing to its half- filled \textit{f} shell with a magnetic moment of S=$\frac{7}{2}$ which makes it more attractive for spintronics applications as compared to the transition metal doped materials. As far as its electronic properties are concerned, it was experimentally demonstrated to be a low carrier semi- metal~\cite{wachkal} and insulating~\cite{xiaochen}. There are also several recent reports~\cite{granville,ludbrook} of GdN having a degenerately doped semiconducting ground state based on the resistivity data measured at low temperatures. Theoretically it is predicted to have a semi- conducting~\cite{lambrecht,ghosh} or a half- metallic character based on \textit{ab- initio} calculations~\cite{aerts,doll}.\\
\indent Initially there had also been a dispute regarding its magnetic properties with earlier observations describing GdN to be a metamagnet~\cite{wachkal,cutlaw} (i.e., at low field an anti- ferromagnetic and at high field a ferromagnetic) material while other studies indicated it to be a ferromagnet~\cite{bus1,schwal,junlevy,mcw,mcg,gambino}. However after such controversial discourse, it has been accepted that GdN is a ferromagnetic~\cite{granville,li,leuenberger} material with experimental reported values of the T$_{\textrm{c}}$ in the range 58 - 90 K~\cite{granville,bus1,schwal,junlevy,mcw,mcg,gambino,li,leuenberger,scarpulla}. But there are as many puzzles on the mechanism of ferromagnetic exchange interaction in GdN as in case of its electronic ground state.\\
\indent This paper is organised as follows. In section~\ref{sec:magprop} we discuss the existing proposed mechanisms behind the magnetic ground state of GdN. In section~\ref{sec:modresdis} we present results of our multiband modified RKKY- like exchange interaction which takes as an input for single particle energies the realistic bandstructure of the 5\textit{d} conduction band. We examine the possibility of carrier- mediated ferromagnetism in GdN. In order to understand the origin of source carriers and their role in supporting ferromagnetism we consider a simple phenomenological model and study its validity. In section~\ref{sec:sumconc} we summarize and conclude our obtained results.\\

\section{\label{sec:magprop} Exchange Interactions}
\indent GdN seem to behave like the ferromagnetic EuO since both have similar magnetic moments and values of T$_{\textrm{c}}$. So one would expect the same mechanism behind exchange interaction in GdN as in EuO. But Kasuya and Li~\cite{kasli1,kasli2} explained the essential distinction which is briefly summarized below.\\ 
\indent There are three known mechanisms for the exchange interactions between magnetic atoms in rare- earth compounds. The first is due to the second order perturbation of the intra- atomic \textit{d-f} exchange giving rise to RKKY~\cite{rkky}- like interaction. The second is due to third order perturbation theory of the \textit{d-f} exchange and \textit{d-f} mixing. The nearest neighbor exchange interaction J$_{1}$ in Europium chalcogenides is due to this mechanism~\cite{maugod}. The third is due to the fourth order perturbation of the \textit{d-f} or \textit{p-f} mixings, where \textit{p} are the anion states~\cite{takkas}. The first mechanism does not depend on the 4\textit{f} level. Whereas the second and third mechanisms become important when the 4\textit{f} level is near the Fermi edge as found in EuO. But in GdN, it is known~\cite{granville,leuenberger} that the 4\textit{f} level lies much below the Fermi level. So the first could be one of the possible mechansims of exchange interaction in GdN. \\
\indent Since both the compounds have different possible mechanisms then what could be the reason for having similar values of T$_{\textrm{c}}$ ?. Let us consider the nearest neighbor exchange interaction in Eu chalcogenides as briefly explained earlier and see how different it is in GdN. \\
\indent The exchange interaction, J$_{1}$ is dominated by an indirect interaction arising from the virtual excitation of a 4\textit{f} (lying inside the semiconducting gap for EuO) to a 5\textit{d} state, which then overlaps the neighboring Eu and leads to a \textit{f}-\textit{f} interaction through the \textit{d-f} exchange. This \textit{d-f} exchange essentially measures the spin splitting of the \textit{d} bands induced by their intra- atomic exchange interaction with the \textit{f} state. One may visualize the effect as arising from the hopping of \textit{f} electron to a neighboring site \textit{d} orbital where it is subject to a spin exchange interaction, J$_{\textrm{df}}$. In the language of perturbation theory it means that the \textit{d} orbital gets mixed into \textit{f} band in an amount $\frac{\textrm{t}_{\textrm{df}}}{(\epsilon_{\textrm{d}}-\epsilon_{\textrm{f}})}$ where $\textrm{t}_{\textrm{df}}$ is the hopping integral and $(\epsilon_{\textrm{d}}-\epsilon_{\textrm{f}})$ is the energy difference between the bottom of the \textit{d} band and the localized \textit{f} level. The contribution to the exchange interaction between the nearest neighbor \textit{f} sites is inversely proportional to $(\epsilon_{\textrm{d}}-\epsilon_{\textrm{f}})$. \\
\indent One would expect a smaller J$_{1}$ in GdN as the above mentioned energy difference is much larger in GdN with the 4\textit{f} level lying several eV below the conduction band edge as compared to EuO where it is inside the semiconducting band gap. Thus J$_{1}$ in accordance with third order perturbation theory is an order of magnitude less in GdN than in EuO. So in order to have similar values of T$_{\textrm{c}}$ in both the compounds there should be \textit{another} dominating indirect exchange mechanism in GdN. \\
\indent Kasuya and Li~\cite{kasli1,kasli2} developed the fourth order perturbation theory which considers the cross process between the \textit{d-f} mixing and \textit{d}-\textit{f} exchange interaction. The resulting effective spin- spin exchange interation depends on inverse of the energy gap i.e., difference between the bottom of the \textit{d} band and the top of the \textit{p} band, $(\epsilon_{\textrm{d}}-\epsilon_{\textrm{p}})$. This exchange interaction energy is evaluated to be large in GdN because the theoretically ascribed energy gap in GdN~\cite{lambrecht} is small as compared to the gap in EuO~\cite{eastman}. But since the experimental nature of the electronic ground state in GdN is not yet clear and the value of the energy gap has not been reported this mechanism remains to be verified. And moreover as mentioned earlier that the 4\textit{f} level in GdN is several eVs below the Fermi level so the mechanisms due to fourth and third order seems less plausible. \\
\indent Recently, Mitra and Lambrecht~\cite{mitra} presented an alternative way to explain the ferromagnetic ground state structure in GdN. There is as an anti- ferromagnetic ordering on a rocksalt lattice between N \textit{p} and Gd \textit{d} magnetic moments. And due to the \textit{d-f} exchange coupling, the nearest neighbor Gd atoms interact ferromagnetically with each other. According to their picture even the next nearest neighbors are ferromagnetically aligned. But they obtained a T$_{\textrm{c}}$ of 10 K within a mean- field calculation which is much lower than observed experimentally. \\
\indent So is there an \textit{additional} indirect exchange mechanism in GdN ? In recent experiments~\cite{ludbrook}, GdN films were found to be semiconducting doped to degeneracy with the most likely source of charge carriers (electrons) as nitrogen vacancies. Although they found the same value of T$_{\textrm{c}}$ as before and also the Hall effect measurements showed the presence of charge carriers but their study with further doping of carriers suggested that exchange is not mediated by free carriers. And this view support the earlier theoretical explanation~\cite{mitra} where the authors abandoned the existence of RKKY- like exchange within rigid band model. \\

\section{\label{sec:modresdis} Models, Results $\&$ Discussion}
\begin{figure}[htbp]
\centering
\vspace*{0.6cm}
\includegraphics[height=6.0cm,width=8.5cm]{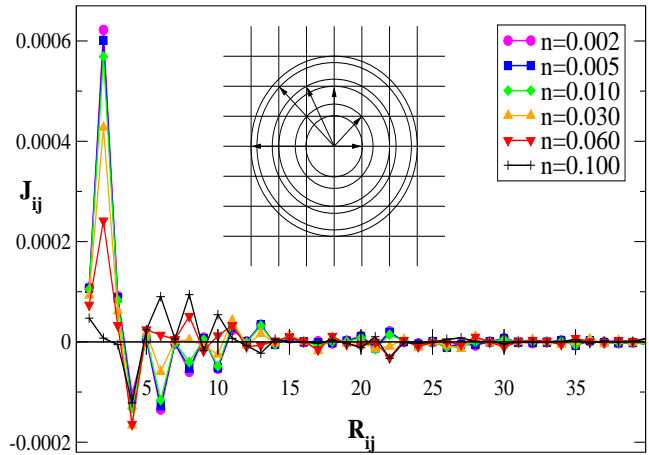}
\vspace*{0.4cm}
\caption{\label{fig:xchint} (Color online) The effective exchange integral as a function of distance for various values of band occupation. The first six shells of nearest neighbors to a central point are shown in a planar geometry.}
\end{figure}

\begin{figure}[ht]
\centering
\vspace*{0.2cm}
\includegraphics[height=6.0cm,width=6.5cm]{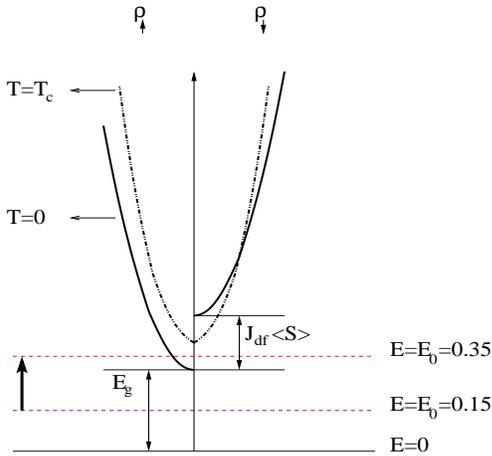}
\caption{\label{fig:resmod} (Color online) A schematic depicting the spin splitting of the density of states below and at the T$_{\textrm{c}}$. A movable impurity level (dashed straight line, E=E$_{\textrm{0}}$) is also shown with a thick arrow pointing its direction.}
\end{figure}

\begin{figure}[ht]
\centering
\includegraphics[height=6.0cm,width=8.5cm]{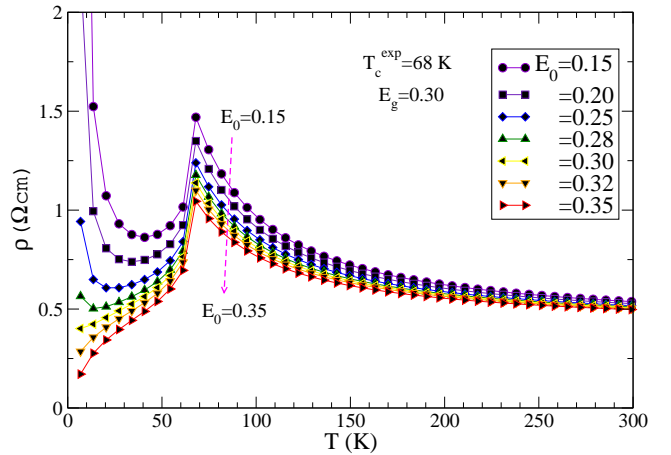}
\caption{\label{fig:r04} (Color online) Temperature dependent resistivity exhibiting metal- to- insulator transition upon the variation of the impurity level, E$_{\textrm{0}}$, out and in the lower edge of conduction band.}
\end{figure}

\indent Here we consider the carrier- mediated ferromagnetism in GdN as a possible \textit{additional} exchange interaction. But our theory is unlike the one studied earlier~\cite{darby,kuznietz}. We are interested in determining the magnetic properties of multiband Kondo lattice model which is described by the following Hamiltonian,
\begin{equation}\label{eq:Ham} 
\textrm{H}=\textrm{H}_{\textrm{kin}}+\textrm{H}_{\textrm{int}}
\end{equation}
where
\begin{equation}\label{eq:Ho}
\textrm{H}_{\textrm{kin}}=\sum_{\textrm{ij}\alpha\beta\sigma}
\textrm{T}_{\textrm{ij}}^{\alpha\beta}\textrm{c}_{\textrm{i}\alpha\sigma}^{\dagger}\textrm{c}_{\textrm{j}\beta\sigma}
\end{equation}
is the kinetic energy of the system and
\begin{equation}\label{eq:Hint}
\textrm{H}_{\textrm{int}}=-\frac{\textrm{J}_{\textrm{df}}}{2}\sum_{\textrm{i}\alpha\sigma}(\textrm{z}_{\sigma}\textrm{S}_{\textrm{i}}^{\textrm{z}}\textrm{c}_{\textrm{i}\alpha\sigma}^{\dagger}\textrm{c}_{i\alpha\sigma}+\textrm{S}_{\textrm{i}}^{\sigma}\textrm{c}_{\textrm{i}\alpha-\sigma}^{\dagger}\textrm{c}_{\textrm{i}\alpha\sigma})
\end{equation}
is the intra- atomic exchange interaction term with an assumption that itinerant electron in each band is coupled to the localized moment by the same coupling strength, J$_\textrm{df}$. The Greek letters ($\alpha$,$\beta$) depict the band indices. In GdN, these are the five \textit{d} conduction bands. The latin letters (i,j) symbolize the crystal lattice sites and spin is denoted by ${\sigma}(={\uparrow},{\downarrow})$. \\
\indent The total Hamiltonian, Eq.~(\ref{eq:Ham}), can be solved using Green function method. Since our interest lies in magnetic properties of multiband Kondo lattice model we have to consider both the sub- systems (localized as well as itinerant) within a self consistent scheme. In Ref.~\cite{sharma1} we have presented our modified RKKY theory which treats both the sub- systems equally. The main idea of the modified RKKY theory is to transform the above Kondo- like exchange Hamiltonian of the conduction electrons into an effective Heisenberg- like spin- spin exchange Hamiltonian of the \textit{f} spins by averaging H$_{\textrm{int}}$ in the subspace of the conduction electrons. In order to avoid repetition of any kind we refer to the reader Ref.~\cite{sharma1} for the complete analysis. But in the following we would like to highlight the numerical details on calculating the exchange integrals and thereby the Curie temperature.\\
\indent The sensitivity of the RKKY- like mechanism to carrier concentration is well known and that the RKKY oscillations are strongly dependent on the value of carrier concentration. The distinct methods~\cite{granville,ludbrook,li,leuenberger,scarpulla,gerlach} of sample preparation lead to different carrier concentrations. This could not only be reason why GdN was noted to be an anti- ferromagnet at low fields with Neel temperature of 40 K~\cite{wachkal,cutlaw} but also why the experimental determination of electronic ground state (semi- metallic or semiconducting) is so uncertain. \\
\indent We take into consideration the semi- conducting nature as predicted theoretically~\cite{lambrecht} and obtained experimentally~\cite{granville,ludbrook}. In the latter case, the carrier concentration (doping) is usually assigned to defects like nitrogen vacancies~\cite{ludbrook} or structural defects (grain boundaries between the nanocrystallites)~\cite{scarpulla}. With realistic values of input parameters~\cite{sharma2} like strength of \textit{d-f} exchange coupling and the single particle energies of 5\textit{d} conduction band obtained using TB-LMTO-ASA~\cite{andersen2} we evaluate the following effective exchange integrals,
\begin{align}\label{eq:exchint}
\textrm{J}^{\textrm{eff}}(\textbf{q}) \nonumber
& = \frac{\textrm{J}_{\textrm{df}}^{2}}{4\pi} \bigg[ Im \int_{-\infty}^{\infty} \textrm{dE} \hspace{0.2cm} \textrm{f}_{-}(\textrm{E}) \frac{1}{\textrm{N}} \sum_{\textrm{ij}\sigma} \widehat{\textrm{G}}_{\textrm{ij}}^{(0)}(\textrm{E}) \widehat{\textrm{G}}_{\textrm{ij}}^{\sigma}(\textrm{E}) \nonumber \\
& \hspace{1.5cm} \textrm{e}^{i\textbf{q} \cdot \big(\textbf{R}_{\textrm{i}} - \textbf{R}_{\textrm{j}}\big)} \bigg] \nonumber \\
& = \sum_{\textrm{ij}} \bigg[ \frac{\textrm{J}_{\textrm{df}}^{2}}{4\pi} Im \int_{-\infty}^{\infty} \textrm{dE} \hspace{0.2cm}  \textrm{f}_{-}(\textrm{E}) \frac{1}{\textrm{N}} \sum_{\sigma} \widehat{\textrm{G}}_{\textrm{ij}}^{(0)}(\textrm{E}) \widehat{\textrm{G}}_{\textrm{ij}}^{\sigma}(\textrm{E}) \bigg] \nonumber \\ 
& \hspace{1.5cm} \textrm{e}^{i\textbf{q} \cdot \big(\textbf{R}_{\textrm{i}} - \textbf{R}_{\textrm{j}}\big)} \nonumber \\
& = \sum_{\textrm{ij}} \textrm{J}_{\textrm{ij}} \textrm{e}^{i\textbf{q} \cdot \big(\textbf{R}_{\textrm{i}} - \textbf{R}_{\textrm{j}}\big)} = \sum_{\textrm{ij}} \textrm{J}_{\textrm{ij}} \textrm{e}^{i\textbf{q} \cdot \big(\textbf{R}_{\textrm{ij}} \big)} \nonumber \\
& = \sum_{\textrm{s},\Delta \textrm{s}} \textrm{J}_{\textrm{s},\Delta \textrm{s}} \textrm{e}^{i\textbf{q} \cdot \textbf{R}_{\textrm{s},\Delta \textrm{s}}} 
\end{align}
where
\begin{equation}\label{eq:jndnmb}
\textrm{J}_{\textrm{s},\Delta \textrm{s}} = \frac{\textrm{J}_{\textrm{df}}^{2}}{4\pi} Im \int_{-\infty}^{\infty} \textrm{dE} \hspace{0.2cm} \textrm{f}_{-}(\textrm{E}) \frac{1}{\textrm{N}} \sum_{\sigma} \widehat{\textrm{G}}_{\textrm{s},\Delta \textrm{s}}^{(0)}(\textrm{E}) \widehat{\textrm{G}}_{\textrm{s},\Delta \textrm{s}}^{\sigma}(\textrm{E})
\end{equation}
\indent In above equations, $\widehat{\textrm{G}}^{(0)}$(E) and $\widehat{\textrm{G}}^{\sigma}$(E) are the single particle non- interacting (undressed) and interacting (dressed) Green function matrices respectively and f$_{-}$(E) is the Fermi function~\cite{sharma1}. The subscript 's' denotes the s$^{\textrm{th}}$ neighboring shell of radius $\textbf{R}_{\textrm{s}}$ spanning $\Delta \textrm{s}$ number of neighbors to the central atom as shown in the inset of Fig.~\ref{fig:xchint} for the case of planar geometry and only for first six shells. But this notation can be generalized to any lattice and finite number of shells until convergence for the exchange integral is reached. \\
\indent In Fig.~\ref{fig:xchint} we also present the dependence of effective exchange integral on the distance for a few low values of carrier concentration. In our numerical calculations we have considered upto 40 shells in 3D geometry of fcc structure for the Gd atoms. As seen in Fig.~\ref{fig:xchint}, the oscillations damp out as a function of distance giving a typical characteristic long range RKKY- like behavior. The atypical part is that the strength of next nearest neighbor interaction is stronger than the nearest neighbor. But it keeps on decreasing rapidly as we increase the carrier concentration and eventually becomes weaker than the nearest neighbor for n=0.1. \\
\indent As the Curie temperature is dependent on effective exchange integrals in the following form~\cite{sharma1},
\begin{equation}\label{eq:curietemp}
\textrm{T}_{\textrm{c}} = \frac{2\textrm{S(S+1)}}{3\textrm{k}_{\textrm{B}}} \bigg[ \frac{1}{\textrm{N}} \sum_{\textbf{q}} \bigg( \frac{1}{ \textrm{J}^{\textrm{eff}}(0) - \textrm{J}^{\textrm{eff}}(\textbf{q}) } \bigg)_{\textrm{T}_{\textrm{c}}} \bigg]^{-1}
\end{equation}
where J$^{\textrm{eff}}(0) = \textrm{J}^{\textrm{eff}}(\textbf{q}=0)$, we calculate T$_{\textrm{c}}$ within a self-consistent scheme for the values of carrier concentrations as considered earlier.\\
\begin{figure}[htbp]
\centering
\vspace*{0.6cm}
\includegraphics[height=6.0cm,width=8.5cm]{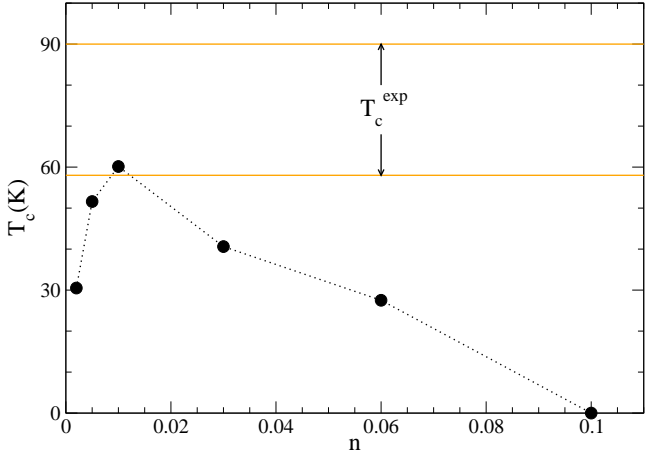}
\vspace*{0.4cm}
\caption{\label{fig:tcvngdn} (Color online) The dependence of T$_{\textrm{c}}$ on band occupation as calculated using modified RKKY theory for J$_{\textrm{df}}$=0.353 eV and S=3.5. The reported experimental range is within 58 - 90 K as shown by two horizontal lines. The dotted black line is a guideline for the eye.}
\end{figure}
\indent Fig.~\ref{fig:tcvngdn} exhibits the dependence of T$_{\textrm{c}}$ on carrier concentration for GdN with the experimentally reported range depicted within the horizontal lines. We obtain our highest value of T$_{\textrm{c}}$=60.144 K for a carrier concentration, n=0.01 ($\sim$ 8 x 10$^{19}$ /cm$^{3}$ for lattice constant of 5 $\AA$). And for concentration of n=0.1 ($\sim$ 8 x 10$^{20}$ /cm$^{3}$) the T$_{\textrm{c}}$ drops down to zero. Such high values of carrier concentration have been observed experimentally~\cite{ludbrook}. \\
\indent In our theory for each value of carrier concentration we determined the Fermi edge self- consistently. We didn't regard any impurity level or band in our calculations. It would be interesting to model the source of charge carriers in a realistic way. And study the contribution of such defects or impurities in stabilizing the carrier- mediated ferromagnetism in GdN. We keep this analysis for future work. But in order to confirm the role of charge carriers and explain one of the anomaly in the low temperature behavior of resistivity~\cite{granville}, we examine a very simple phenomenological model as shown in Figure~\ref{fig:resmod}.\\
\indent We consider the spin splitting of the parabolic model density of states below T$_{\textrm{c}}$ within the mean field picture having energy difference of the order $\sim$ J$_{\textrm{df}}$S where S is the 4\textit{f} local moment. An impurity level which can be moved in or out of the spin-up bands is also shown. Although recent experiments~\cite{ludbrook} support the presence of charge carriers (electrons) due to impurities but no evidence is reported on whether do these carriers form an impurity level or degenerate impurity band. We take into account an impurity level instead of a degenerate band. It is known that the impurity concentration in case of intrinsic semiconductor at room temperature is proportional to e$^{\frac{-\Delta \textrm{E}}{\textrm{k}_{\textrm{B}}\textrm{T}}}$. So we consider the resistivity to be proportional to e$^{\frac{\Delta \textrm{E}}{\textrm{k}_{\textrm{B}}\textrm{T}}}$ and is given as,
\begin{equation}\label{eq:resist}
\rho = \rho(\textrm{0}) \hspace*{0.2cm} \textrm{e}^{\frac{\Delta \textrm{E}}{\textrm{k}_{\textrm{B}}\textrm{T}}}
\end{equation}
where the activation energy, $\Delta \textrm{E}$, is given by 
\begin{equation}\label{eq:resformula}
\Delta \textrm{E} = (\textrm{E}_{\textrm{g}} - \textrm{E}_{0}) + \frac{\textrm{J}_{\textrm{df}}\textrm{S}}{2}\bigg(1 - \frac{\langle \textrm{S}^{\textrm{z}} \rangle}{\textrm{S}}\bigg)
\end{equation}
\indent The reason to assume such a form of activation energy can be understood as follows. Our goal is to model a ferromagnetic semiconductor with a finite gap E$_{\textrm{g}}$ $\gg$ k$_{\textrm{B}}$T and described within Kondo lattice model. It governs the temperature dependence via the magnetization ($\langle \textrm{S}^{\textrm{z}} \rangle$) which we consider within the molecular field theory. Below the ferromagnetic T$_{\textrm{c}}$ the resistivity depends on the scattering of charge carriers due to their interaction with localized moments. Above T$_{\textrm{c}}$, the resistivity follows the normal thermally activated energy behavior and falls off exponentially with further increase in the temperature.\\
\indent The first term in the bracket of Eq.~(\ref{eq:resformula}) represents the loss in binding (trap) energy of the electrons from the impurity level and the second term is the exchange contribution that reaches its maximum value at or above T$_{\textrm{c}}$. Using the parameters from the literature~\cite{granville,lambrecht,sharma2} i.e. J$_\textrm{df}$=0.353 eV, S=$\frac{7}{2}$ , E$_{\textrm{g}}$=0.30 eV, T$_{\textrm{c}}^{\textrm{exp}}$ = 68 K and $\rho(\textrm{0})$=0.4 $\Omega$cm, we try to evaluate Eq.~(\ref{eq:resformula}) for different values of E$_{\textrm{0}}$. We move the impurity level from the center of the gap into the lower edge of the spin- up density of states of the conduction band. The result is shown in Figure~\ref{fig:r04}.\\
\indent Our simple model demonstrates an important effect like insulator- to- metal transition at low temperature by moving the level E$_{\textrm{0}}$ as shown in Figure~\ref{fig:resmod}. Similar effect induced by applying strain was theoretically reported based on first principles calculations~\cite{duan2}. Thus it clearly reveals that the experimentally observed resistivity~\cite{granville} is an artifact of a degenerate impurity level lying close to the lower edge of \textit{d} conduction band. And carriers in such level play an essential (\textit{additional}) part in stiffening the ferromagnetism in GdN.\\

\section{\label{sec:sumconc} Summary $\&$ Conclusion}
\indent In conclusion the basic exchange mechanism in GdN is not clear. There is a theory based on fourth order perturbation~\cite{kasli1} which seem less probable since it requires the 4\textit{f} level to be near the Fermi edge while experimentally it is known to be several eVs below. There is another~\cite{mitra} interpretation but it results in obtaining much lower T$_{\textrm{c}}$ as compared to the experimentally reported values. \\
\indent We consider our multiband modified RKKY theory where an effective spin Hamiltonian is obtained by integrating out charge degress of freedom from the multiband Kondo lattice model~\cite{sharma1}. We take the realistic bandstructure of 5\textit{d} conduction bands as an input for the single particle energies and the \textit{d-f} exchange coupling~\cite{sharma2} to calculate the dependence of T$_{\textrm{c}}$ on carrier concentration. The results are in close proximity to the experimental findings. \\
\indent In order to trace the source of charge carriers which are eventually responsible for an \textit{additional} carrier- mediated ferromagnetism in GdN we consider a simple phenomenological model. It not only explains experimentally observed anomaly in the low temperature behavior of resistivity~\cite{granville} but also exhibits insulator- to- metal transition in accordance with theory~\cite{duan2}. Our results indicate that if pure (stoichiometric) GdN is prepared it will have a low T$_{\textrm{c}}$ as predicted earlier~\cite{mitra}. \\

\section{Acknowledgement}
\indent We would like to thank Prof. W. Borgiel for his critical reading of the manuscript and useful comments.

\end{document}